\documentstyle[12pt]{article}
\begin{document}
\title{A NOTE ON THE PAPER "QUARKS OR LEPTONS?"}
\author{B.G. Sidharth$^*$\\ Centre for Applicable Mathematics \& Computer Sciences\\
B.M. Birla Science Centre, Hyderabad 500 063 (India)}
\date{}
\maketitle
\footnotetext{$^*$E-mail:birlasc@hd1.vsnl.net.in}
\begin{abstract}
In the context of a recent description of Fermions as Kerr-Newman type
black holes with Quantum Mechanical inputs, it is shown how the quark
picture can be recovered. The advantage is that in the process we obtain
a rationale for such features as the puzzling fractional charges of the
quarks, their masses, confinement and handedness in a unified scheme.
\end{abstract}
In a previous paper\cite{r1}, it was suggested in the spirit of Barut, that
Leptons alone could be considered to be fundamental, and it was attempted to
explain the proton structure in this light.\\
In the present note we examine the above considerations in the light of a recent
description\cite{r2,r3,r4} of Fermions as Kerr-Newman type Black Holes in
a Quantum Mechanical setting and show that we can equivalently recover the
usual quark picture. The advantage is that in the process we obtain a rationale for such features
as the puzzling fractional charges of the quarks, their masses, confinement
and handedness in a unified scheme.\\
Indeed it is well known that the Kerr-Newman metric describes the field of an
electron including the anomalous gyro magnetic ratio, $g = 2.$ However there
appears a naked singularity which in the above description disappears due to
Quantum Mechanical Zitterbewegung effects within the Compton wavelength. This
description was shown to provide a rationale for several hitherto empirical
features like the discreteness of the charge and the handedness of the
neutrino, while at the same time providing a reconciliation between gravitation
and electromagnetism.\\
It was shown (cf.ref.\cite{r2} and\cite{r3})
that near the Compton wavelength the potential is given by
\begin{eqnarray}
4m \int \frac{T_{\mu \nu} (t,\vec x')}{|\vec x - \vec x' |} d^3 x' +
(\mbox terms \quad independent \quad of \quad \vec x), \nonumber \\
+ 2m \int \frac{d^2}{dt^2} T_{\mu \nu} (t,\vec x')\cdot |\vec x - \vec x' |
d^3 x' + 0 (| \vec x - \vec x' |^2)\label{e1}
\end{eqnarray}
Outside the Compton wavelength, it was shown that this leads to
\begin{eqnarray}
\frac{ee'}{r} = A_0 \approx \frac{2c \hbar}{r} \int \eta^{\mu \nu} \frac{d}{d \tau}
T_{\mu \nu} d^3 x' = \frac{2c \hbar}{r} \int \eta^{\imath j} \frac{d}{d \tau}
T_{\imath j} d^3 x', \nonumber \\
=  2c \hbar (\frac{mc^2}{\hbar}) \int \eta^{\imath j}
\frac{T_{\imath j}}{r} d^3 x'\label{e2}
\end{eqnarray}
where $e'$ is the test charge.\\
As we approach the Compton wavelength however, we have to use equation
(\ref{e1}), which after a division by $m$, the mass of the particle
to be identified with the quark, and taking $\hbar = 1 = c$ to correspond to
the usual theory\cite{r5}, goes over to
\begin{equation}
-\frac{\alpha}{r} + \frac{\beta m_e}{l^2}r\label{e3}
\end{equation}
where $\alpha \sim 1$ and $\beta \sim \frac{1}{m}$ and $m_e$ is the electron
mass. It was pointed out for example in\cite{r3} that this is the QCD potential with both the Coulumbic and confining
parts.\\
We now observe that the usual three dimensionality of space, as pointed out
by Wheeler\cite{r6} arises due to the doubleconnectivity or spinorial
behaviour of Fermions, which takes place outside
the Compton wavelength due to the fact that while it is the negative
energy components of the Dirac four-spinor which dominate inside, it is the
positive energy components which predominate outside (cf.ref.\cite{r2}to\cite{r4} for
details). But as we approach the Compton wavelength, we encounter mostly the
negative energy components and the above doubleconnectivity and therefore
three dimensionality disappear: We have two or less dimensions. Indeed such
a conclusion has been drawn alternatively at very small scales (cf.\cite{r7,r8}).
Further recent experiments with nano tubes already
reveal such low dimensional quantum behaviour\cite{r9,r10}.\\
This leads to the following circumstance:
We first have to consider two and one spatial dimensions.
We now use the fact that as is well known\cite{r11}
each of the $T_{\imath j}$ in (\ref{e2}) is given by $(1/3) \epsilon$, where
$\epsilon$ is the energy density. In this case it follows from (\ref{e2})
that the particle would have the charge $(2/3) e$ or $(1/3) e$, in two or one
dimensions. Incidentally, this provides an explanation for the remarkable
and well known fact that a third of charge appears to be concentrated in
a core of the size of the order of the Compton wavelength\cite{r12}.\\
This would also automatically imply that these fractionally charged particles
cannot be observed individually, as they are by their very nature confined to dimensions of the
order of their Compton wavelength. This is expressed by the confining
part of the QCD potential (\ref{e3}). We now identify these confined particles
with charge $(1/3) e$ and $(2/3) e$, with quarks and further justify this identification,
below.\\
As in reference\cite{r1}, and as in the standard theory we consider the proton to be
made up of two quarks of charge $(2/3) e$ with an intervening quark of charge
$-(1/3) e$ all confined to a distance $l$ which in the above light is of the
order of the particle's Compton wavelength. For small displacements $r$ of the
central quark as in\cite{r1} we can easily see that the confining part of the potential is
given by
\begin{equation}
V = \frac{e^2}{9l^2}\quad r\label{e4}
\end{equation}
Comparing with (\ref{e3}) we get,
$$\frac{e^2}{9l^2} \quad \sim \frac{1}{m} \quad \frac{m_e}{l^2}$$,
whence the quark's mass is given by
\begin{equation}
m \sim 10^3 m_e\label{e5}
\end{equation}
as required.\\
Finally as we encounter predominantly the negative energy two spinor of the
Dirac four spinor at the Compton wavelength, with negative helicity
(cf.ref.\cite{r3}and \cite{r13}), the quarks display handedness which in
conventional theory is due to the small Cabibo angle.\\
It is interesting to
note that just beyond the Compton wavelength, where we still do not
encounter fractional charge or low dimensions, the mass of the resulting
particle would from (\ref{e4}) and (\ref{e5}) be given by $\sim 137 m_e$
corresponding to the pion: In fact we have to consider two such fermions,
as is well known, so that we recover the pion mass, $274 m_e$.\\
In other words at scales greater than the Compton wavelength the above
description will correspond to that of an electron, while at scales $\ge_\sim$
the Compton wavelength, it corresponds to a pion and at scales $\le_\sim$ the
Compton wavelength, it corresponds to a quark, which is physically
meaningful.\\
In any case we recover the usual structure of the proton in terms of the
three quarks.

\end{document}